\documentclass[aps,prd,twocolumn,reprint,amsmath,amssymb,preprintnumbers,superscriptaddress,nobibnotes,nofootinbib,floatfix]{revtex4-1}

\usepackage{graphicx}
\usepackage{dcolumn}
\usepackage{bm}
\usepackage{color}
\usepackage{amssymb}
\usepackage{amsmath}
\usepackage[colorlinks=true,allcolors=Purple,pdfusetitle]{hyperref}
\usepackage{rotating}
\usepackage{multirow}
\usepackage{graphicx}
\usepackage[dvipsnames]{xcolor}
\usepackage{svg}
\usepackage{esdiff}
\definecolor{Green}{RGB}{0, 128, 0}
\newcommand{\orcid}[1]{\href{https://orcid.org/#1}{\includegraphics[width=10pt]{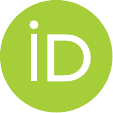}}}
\makeatletter

\begin{document}
\preprint{FERMILAB-PUB-24-0845-T, N3AS-24-038}

\title{Searching for MeV-mass neutrinophilic Dark Matter with Large Scale Dark Matter Detectors}

\author{Anna M. Suliga\orcid{0000-0002-8354-012X}}
\email{a.suliga@nyu.edu}
\affiliation{Center for Cosmology and Particle Physics, New York University, New York, NY 10003, USA}
\affiliation{
Department of Physics, University of California Berkeley, Berkeley, California 94720, USA}
\affiliation{
Department of Physics, University of California, San Diego, La Jolla, CA 92093-0319, USA}

\author{George~M.~Fuller\orcid{0000-0002-4203-4108}}
\email{gfuller@physics.ucsd.edu}
\affiliation{
Department of Physics, University of California, San Diego, La Jolla, CA 92093-0319, USA}

\date{\today}

\begin{abstract}
The indirect detection of dark matter (DM) through its annihilation products is one of the primary strategies for DM detection. One of the least constrained classes of models is neutrinophilic DM, because the annihilation products, weakly interacting neutrinos, are challenging to observe. 
Here, we consider a scenario where MeV-mass DM exclusively annihilates to the third neutrino mass eigenstate, which is predominantly of tau and muon flavor. In such a scenario, the potential detection rate of the neutrinos originating from the DM annihilation in our Galaxy in the conventional detectors would be suppressed by up to approximately two orders of magnitude. This is because the best sensitivity of such detectors for neutrinos with energies below approximately 100~MeV is for electron neutrino flavor. In this work, we highlight the potential of large-scale DM detectors in uncovering such signals in the tens of MeV range of DM masses. In addition, we discuss how coincident signals in direct detection DM experiments and upcoming neutrino detectors such as DUNE, Hyper-Kamiokande, and JUNO could provide new perspectives on the DM problem.
\end{abstract}
\maketitle

\section{Introduction}
\label{sec:Intro}

Despite the overwhelming evidence for the existence of Dark Matter (DM) due to its impact on the cosmological and astrophysical environments by its gravitational effects~\cite{1980ApJ...238..471R, 1987ARA&A..25..425T, Planck:2018vyg}, little is known about its properties and exact nature~\cite{Bertone:2004pz, Slatyer:2017sev}. 
The lack of measurable signals over a wide range of energies in the terrestrial and astrophysical DM searches puts stringent constraints on nongravitational DM interactions over a broad range of masses~\cite{Essig:2013goa, Slatyer:2015jla, Slatyer:2017sev, Leane:2018kjk, Dutta:2022wdi, Boddy:2022knd, Baryakhtar:2022hbu, Cooley:2022ufh, Aalbers:2022dzr, Steigman:2012nb, Beacom:2006tt, Yuksel:2007ac, Palomares-Ruiz:2007trf, KamLAND:2011bnd, Kappl:2011kz, Frankiewicz:2015zma, IceCube:2015rnn, Albert:2016emp, Frankiewicz:2017trk, Hiroshima:2017hmy, Olivares-DelCampo:2017feq, Klop:2018ltd, Arguelles:2019ouk, Bell:2020rkw, Ferrer:2022kei, Cline:2022qld, Fujiwara:2023lsv, McKeen:2023ztq, Das:2024bed}. 
If DM annihilates into ionizing particles, strong limits exist on the annihilation cross-sections. Planck~\cite{Planck:2015fie, Planck:2018vyg} measurements of of the cosmic microwave background (CMB) radiation can rule out thermal DM in the form of Weakly Interacting Massive Particles (WIMPs) with rest masses below approximately 20~GeV because these particles could affect the CMB anisotropies that are measured to high precision by Planck. They would do this by changing the ionization history of the hydrogen and helium in the early Universe~\cite{Slatyer:2015jla}. Such annihilation scenarios would also change the astrophysical fluxes measured by various telescopes and could appear in the collider searches, see, e.g., Refs.~\cite{Essig:2013goa, Slatyer:2017sev, Dutta:2022wdi} for recent summaries on the DM annihilation limits. 

The least stringent bounds are on the interactions of DM with the most feebly interacting Standard Model particles, e.g., neutrinos, or potential dark sector particles~\cite{Steigman:2012nb, Beacom:2006tt, Yuksel:2007ac, Palomares-Ruiz:2007trf, KamLAND:2011bnd, Kappl:2011kz, Murase:2012xs, Frankiewicz:2015zma, IceCube:2015rnn, Albert:2016emp, Frankiewicz:2017trk, Hiroshima:2017hmy, Olivares-DelCampo:2017feq, Olivares-DelCampo:2018pdl, Klop:2018ltd, Hyper-Kamiokande:2018ofw, Arguelles:2019ouk, Bell:2020rkw, Okawa:2020jea, Ferrer:2022kei, Cline:2022qld, Fujiwara:2023lsv, McKeen:2023ztq, Pospelov:2023mlz, Das:2024bed}, see Ref.~\cite{Arguelles:2019ouk} for a recent review. When thermally produced DM particles annihilate exclusively to neutrinos, the DM mass can be as low as $\mathcal{O}(5-10)$\;MeV. Lower DM masses would affect the effective number of neutrinos encoded in the CMB and Big Bang Nucleosynthesis probes~\cite{Boehm:2013jpa, Nollett:2013pwa, Nollett:2014lwa, Escudero:2018mvt, Sabti:2019mhn, Wang:2023csv, Chu:2023jyb, Chu:2023zbo}.

\begin{figure}[t]
\includegraphics[width=0.99\columnwidth]{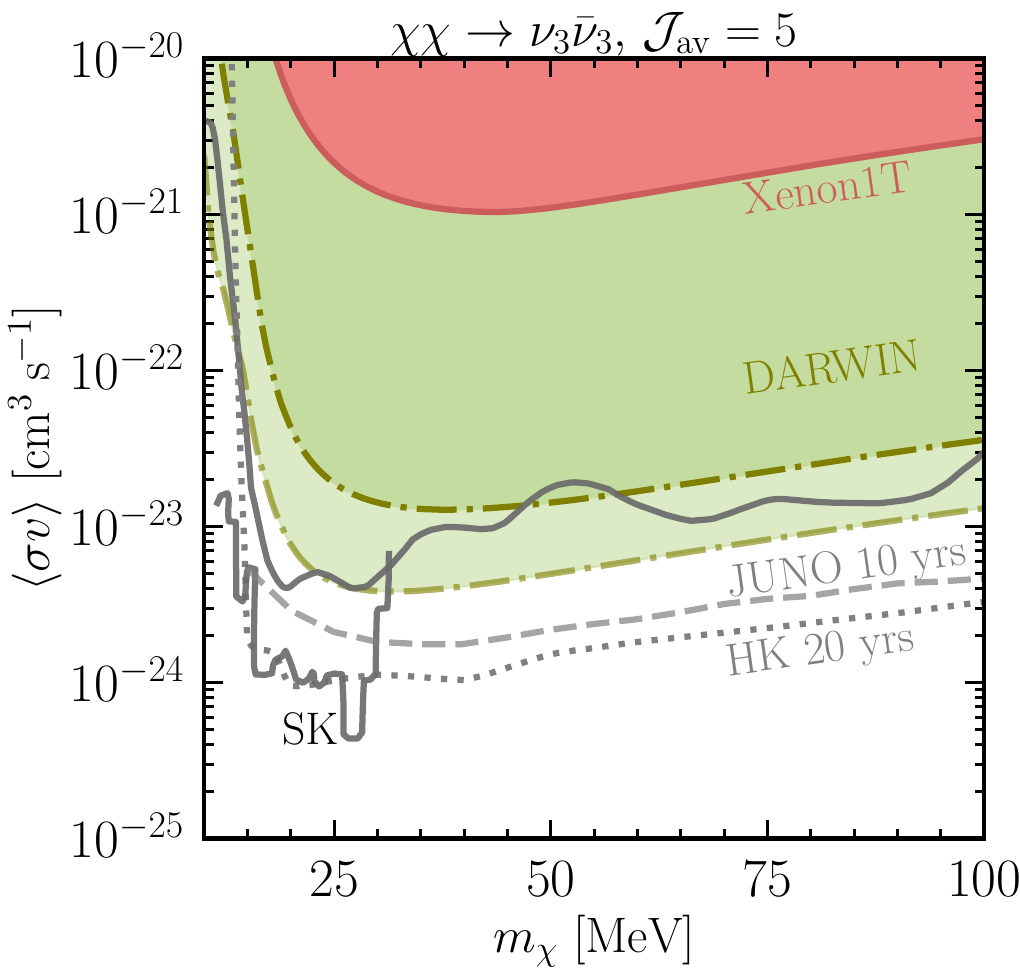}
\caption{The 90\% C.L. sensitivity limits on the DM particle $\chi$ annihilation cross section to $\nu_3$ neutrino mass eigenstate in our Galaxy from Xenon1T (red region) and DARWIN (green regions with dashed-dotted border lines, see text for details) together with the re-casted limits from SK (gray solid line)~\cite{Olivares-DelCampo:2017feq, Arguelles:2019ouk} and sensitivity limits from HK~\cite{Olivares-DelCampo:2018pdl} (gray dotted line), and JUNO~\cite{JUNO:2023vyz} (gray dashed line).}  
\label{Fig1}
\end{figure} 

Here we explore a scenario where the DM particles, with masses in the range of 10 - 100\;MeV, annihilate to the neutrino mass state containing the least electron flavor, i.e, $\nu_3$. For models that could potentially accommodate this scenario, see, e.g., Refs.~\cite{Boehm:2006mi, Pospelov:2007mp, Bertoni:2014mva, Garcia-Cely:2017oco, Batell:2017cmf, Herms:2023cyy, Xu:2023xva, Akita:2023qiz, Jana:2024iig, Cardenas:2024ojd}.
For example, for the majoron or scotogenic like models the branching ratio for DM annihilation to neutrinos scales as mass squared. In the normal mass hierarchy~\cite{DESI:2024mwx}, in these models the annihilation probability to the second mass eigenstate is approximately 30 times smaller than that into the third mass eigenstate for the atmospheric neutrino mass squared difference. With the first mass eigenstate having a mass close to zero, this results in DM effectively annihilating to $\nu_3$. The contribution of the electron flavor to that mass state is approximately $|U_{e3}|^2 \approx 2\%$~\cite{ParticleDataGroup:2024cfk}.

The salient feature of this assumption is that the electron neutrino ($\nu_e$) component of the flux created by the DM annihilators remains low. This is not the case when the DM annihilates primarily to $\nu_e$, all flavors equally, or even $\nu_\mu$ and $\nu_\tau$ flavors states only. Assuming that the neutrino mass eigenstate decohere on their way to the Earth, and only vacuum oscillations take place, the conversion probability for $P_{\nu_\tau \rightarrow \nu_e} = \sum_i |U_{e, i}|^2 |U_{\tau, i}|^2 \approx 0.23$ and $P_{\nu_\mu \rightarrow \nu_e} \approx 0.22$, where $U_{\alpha,i}$ are the entries of the Pontecorvo-Maki-Nakagawa-Sakata (PMNS) matrix~\cite{Pontecorvo:1957qd, Maki:1962mu} and using the values for the mixing angles $\theta_{12} = 33.41$, $\theta_{23}=49.10$, and $\theta_{13} = 8.54$~\cite{ParticleDataGroup:2024cfk}. 
The final flavor ratios for the initial pure $\nu_\tau$ or $\nu_\mu$ fluxes due to the neutrino oscillations taking place as they are traveling to Earth in those cases would be $\nu_e$ : $\nu_\mu$ : $\nu_\tau \approx$ 1 : 1.7 : 1.8 or 1: 1.8 : 1.7, i.e, approximately 23 or 22\% of the initial $\nu_\tau$ or $\nu_\mu$ flux would convert to electron flavor neutrinos. Additionally, for any scenario where DM annihilates to neutrino flavor states with an approximately democratic flavor ratio, the appearance of a large $\nu_e$ component due to neutrino oscillations is inevitable. For example, for an initial 1 : 1 : 1 ratio, the final ratio would also include a 1/3 $\nu_e$ component. For DM annihilation to $\nu_e$ only, the final flavor ratio is 2.5 : 1 : 1, i.e., $P_{\nu_e\rightarrow\nu_e} = 0.55$ of the initial $\nu_e$ flux survives.

On the contrary, in our scenario --- DM annihilates exclusively to $\nu_3$ --- a significantly large non-electron component can be created by the DM annihilation, while the electron neutrino component remains low, i.e., $|U_{e3}|^2 \approx 0.02$~\cite{ParticleDataGroup:2024cfk}. Any potential medium-induced re-generation of $\nu_e$ stemming from matter effects inside Earth will be negligibly small~\cite{Dighe:1999bi}.

Detecting low-energy, i.e., 10 - 100\;MeV, non-electron neutrino fluxes is challenging in traditional neutrino detectors. The muon and tau neutrinos can only be observed through their neutral-current interactions when these neutrinos have energies below their corresponding charged lepton mass thresholds. These interactions have smaller cross sections than the charged-current ones available for the electron flavor neutrinos. In addition, coincidence detection is not achievable in those channels because only a single interaction product is observed, making the background rejection a formidable task.

The interaction channel that has a large cross section for the neutral-current neutrino interactions in the tens of MeV neutrino energy range is Coherent Elastic Neutrino-Nucleus Scattering (CE$\nu$NS)~\cite{Freedman:1973yd}. That reaction has been observed by the COHERENT experiment~\cite{COHERENT:2017ipa}. 
For non-terrestrial neutrino fluxes the COHERENT experiment's detector volume is too small to observe them with CE$\nu$NS. However, as the DM detectors become more and more sensitive~\cite{EDELWEISS:2011epn, 2012EPJC...72.1971A, TEXONO:2013hrh, DEAP-3600:2017uua, DarkSide:2018bpj, SuperCDMS:2018mne, LUX:2018akb, XENON:2018voc, PandaX-4T:2021bab}, they can ultimately detect the so-called neutrino floor (or fog)~\cite{Vergados:2008jp, Strigari:2009bq, Billard:2013qya, Baudis:2013qla, Ruppin:2014bra, OHare:2016pjy, Boehm:2018sux}. 
Currently, XENONnT~\cite{XENON:2024ijk} and PandaX-4T~\cite{PandaX:2024muv} infer an approximately 2.7$\sigma$ preference for a non-zero solar neutrino flux. 

In this work, we utilize both the advantages of the large CE$\nu$NS cross section and large volume xenon-based DM detectors to calculate the sensitivity limits to the annihilation cross section for the DM annihilating to $\nu_3$. The advantages of the large scale DM detectors for probing DM annihilation producing a $\nu_\tau$ flux also have been studied in Ref.~\cite{McKeen:2018pbb}, where the possibility of decays to mass states was discussed too. The advantages of these direct DM detectors for the boosted DM scenarios also have been pointed out in Refs.~\cite{Cherry:2015oca, Das:2024ghw, Ghosh:2024dqw}.

Figure~\ref{Fig1} shows the parameter space region for the DM annihilating to $\nu_3$ that can be probed with the past, existing, and planned large-scale DM detectors such as Xenon1T~\cite{XENON:2017lvq}, PandaX-4T~\cite{PandaX-4T:2021bab}, XENONnT~\cite{XENON:2020kmp}, LUX-ZEPLIN~\cite{LZ:2019sgr} and DARWIN~\cite{DARWIN:2016hyl}. The most stringent existing limits on the scenrio with DM annihilation to $\nu_3$ in this range of DM masses comes from the Super-Kamiokande (SK) experiment~\cite{Palomares-Ruiz:2007trf, Frankiewicz:2015zma, Frankiewicz:2017trk,  Olivares-DelCampo:2017feq, Arguelles:2019ouk}. 
Here, in this work, we find a potential discovery region for this scenario, a non-excluded region of the parameter space that can be probed by both the neutrino and DM detectors such as Hyper-Kamiokande (HK)~\cite{Olivares-DelCampo:2018pdl, Bell:2020rkw} and JUNO~\cite{Akita:2022lit, JUNO:2023vyz}.  
We re-cast the limits from SK~\cite{Olivares-DelCampo:2017feq}, HK~\cite{Olivares-DelCampo:2018pdl} and JUNO~\cite{JUNO:2023vyz} on $\chi\chi \rightarrow \sum_\alpha \nu_\alpha \bar\nu_\alpha$ to $\chi\chi \rightarrow \nu_3\bar\nu_3$ by multiplying them by a factor 1/3 and dividing by $|U_{e3}|^2$ to account for the oscillations of electron neutrinos to in flight and the fraction of the electron neutrino in the third mass eigenstate.
The limits from the HK analysis in~\cite{Bell:2020rkw} to $\chi\chi \rightarrow \nu_e \bar\nu_e$ need to be recast to $\chi\chi \rightarrow \nu_3\bar\nu_3$ by multiplying them by a factor 0.55 and dividing by $|U_{e3}|^2$.

The rest of this manuscript is organized as follows. In Sec.~\ref{sec:Flux}, we review how to calculate the flux of neutrinos coming from the DM annihilation in our Galaxy in an approximate way. Following, in Sec~\ref{sec:Limits}, we first calculate the expected event rates in the xenon-based DM detectors from DM annihilation to $\nu_3$ (Sec.~\ref{sec:EventRates}), and then we show our sensitivity limits calculated for XENON1T and DARIWN (Sec.~\ref{sec:SensitivityLimits}). We discuss and summarize our findings in Sec.~\ref{sec:Discussion_and_Conclusions}.

\section{Modeling the neutrino flux from Dark Matter annihilation}
\label{sec:Flux}

We model the neutrino energy $E_\nu$ dependent flux of the $\nu_3$ and $\bar\nu_3$ mass states created by the Majorana type DM annihilation averaged over the Milky Way galactic halo following the approximation from Ref.~\cite{Palomares-Ruiz:2007trf}
\begin{equation}
\label{eq:flux}
    \frac{dN_{\nu}}{dE_\nu} = \mathcal{J}_\mathrm{av} \frac{\langle \sigma v  \rangle}{2} {R_*}\frac{\rho_*^2}{m_\chi^2} \delta(E_\nu - m_\chi) \ ,
\end{equation}
where $\mathcal{J}_\mathrm{av}$ is the so-called average J-factor, which takes into account the contribution to the flux from the angular-averaged DM distribution over the whole Milky Way halo~\cite{Yuksel:2007ac}. The DM velocity averaged annihilation cross section is $\langle \sigma v  \rangle$, $R_* = 8.5\;\mathrm{kpc}$ is the distance to the Galactic Center, $\rho_\star=0.3\;\mathrm{GeV/cm}^{3}$ is the DM density in the solar circle, $m_\chi$ is the mass of the annihilating DM particle, and $\delta$ is the delta function.
We note that we are using a canonical estimation of the average J-factor, $J_{av} = 5$~\cite{Yuksel:2007ac}, which was also used for the JUNO~\cite{JUNO:2023vyz}, SK~\cite{Olivares-DelCampo:2017feq}, and HK~\cite{Olivares-DelCampo:2018pdl} limits calculations therefore the comparison with the limits from these detectors is straightforward.


\section{Sensitivity limits from Dark Matter Direct Detection Experiments}
\label{sec:Limits}

In this section, we first review how to calculate the CE$\nu$NS event rates in the large scale xenon-based DM detectors~\ref{sec:EventRates}. Following, in Sec.~\ref{sec:SensitivityLimits}, we show our calculated sensitivity limits. These are presented in Fig.~\ref{Fig1}. 

\subsection{Event Rates}
\label{sec:EventRates}
The expected CE$\nu$NS event rates, resulting from flux of $\nu_3$ neutrinos created by the DM annihilation in our Galaxy are 
\begin{equation}
\label{eq:Rate}
    \frac{dR_\nu}{dE_r} \approx N_N \Delta T \epsilon \!\int dE_\nu \frac{d\sigma_{\nu N}}{dE_r} \frac{dN_{\nu}}{dE_\nu} \Theta(E_r^\mathrm{max}-E_r) \ ,
\end{equation}
where $N_N$ is the number of the nuclei for CE$\nu$NS, $\Delta T$ is the exposure time, and where necessary, $\epsilon$ is the $E_r$ -- recoil energy -- dependent detector efficiency. Here $\Theta$ is the Heaviside step function, and $E_r^\mathrm{max} = 2E_\nu^2/(m_N + 2E_\nu)$ is the maximum recoil energy for a given target nucleus with the interacting neutrino energy $E_\nu$. The differential CE$\nu$NS cross-section for a neutrino with energy $E_\nu$ interacting with a heavy nucleus $N$ is given by~\cite{Freedman:1973yd}
\begin{equation} 
\label{cross-section}
    \frac{d\sigma_{{\nu} N}}{dE_r} \simeq \frac{G_\mathrm{F}^2 m_N}{4\pi} Q_W^2\left(1- \frac{m_N E_r}{2E_{\nu}^2}\right)F^2(Q)~,
\end{equation}
where, following common practice, we take $m_N \approx 931.5 A$~MeV as the mass of the target nucleus $N$ with $A$ being its mass number. Here $Q_W = [N - Z(1 - 4$sin$^2(\theta_W)]$ is the weak vector nuclear charge, with sin$^2(\theta_W) = 0.23863$ being the sine-squared of the Weinberg angle at low-energy momentum transfer~\cite{ParticleDataGroup:2022pth}. 
The coherence of the interaction depends on the amount of the momentum transferred to the nucleus $Q = \sqrt{2m_N E_r}$. Here, $F(Q)$ is the form factor taken to be the Helm form factor~\cite{Helm:1956zz}
\begin{equation} \label{form_factor}
    F(Q) = 3\frac{j_1(Q R_0)}{Q R_0} \exp\left(-\frac{1}{2} Q^2 s^2 \right) \ ,
\end{equation}
where $j_1$ is the first Bessel function, the size of the nucleus $N$ undergoing CE$\nu$NS is given by $R_0=\sqrt{R^2-5s^2}$ with $R=1.2A^{{1}/{3}}$ and the nuclear skin thickness of the nucleus $N$ is assumed to be $s \simeq 0.5$.

\begin{figure}[t]
\includegraphics[width=0.99\columnwidth]{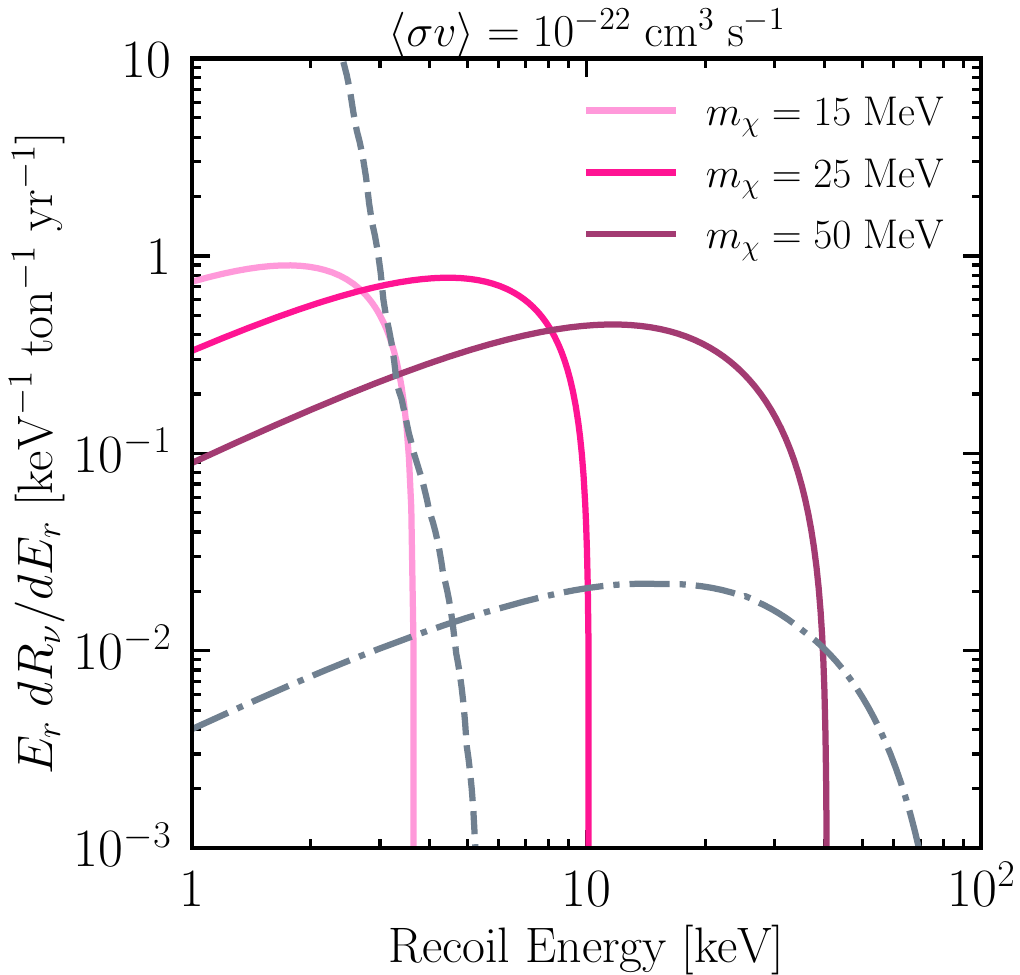}
\caption{Ideal rates of the $\nu_3$ fluxes created by the DM annihilation in our galaxy for $\langle \sigma v \rangle = 10^{-22}\;\mathrm{cm}^3\;\mathrm{s}^{-1}$ and $m_\chi=15\;\mathrm{MeV}$ (solid light pink line), $m_\chi=25\;\mathrm{MeV}$ (solid pink line), and  $m_\chi=50\;\mathrm{MeV}$ (solid dark pink line) together with the sources of irreducible backgrounds, i.e., solar neutrinos (dashed gray line) and atmospheric neutrinos (dash-dotted gray line).} 
\label{Fig2}
\end{figure} 

Figure~\ref{Fig2} shows the ideal nuclear recoil event rates for the fluxes of $\nu_3$ created by the DM annihilation for $\langle \sigma v \rangle =  10^{-22}\;\mathrm{cm}^3\;\mathrm{s}^{-1}$ and $m_\chi=15\;\mathrm{MeV}$ (solid light pink line), $m_\chi=25\;\mathrm{MeV}$ (solid pink line), and  $m_\chi=50\;\mathrm{MeV}$ (solid dark pink line) together with the sources of irreducible backgrounds, i.e., solar neutrinos~\cite{Vinyoles:2016djt, Vitagliano:2019yzm} (dashed gray line) and atmospheric neutrinos~\cite{Battistoni:2002ew, Newstead:2020fie} (dash-dotted gray line). The large flux of solar neutrinos below recoil energies of approximately 3~keV significantly suppresses the sensitivity to DM particles with masses below $\lesssim \;15\;\mathrm{MeV}$. We neglect the contributions from the diffuse supernova neutrino background~\cite{Ando:2004hc, Lunardini:2010ab, Beacom:2010kk, Ando:2023fcc, Suliga:2022ica} to the irreducible background for this search because it is expected to be much smaller than the solar and atmospheric neutrino contributions~\cite{Suliga:2021hek, Zhuang:2024exm}, unless there are unknown effects that increase the non-electron part of that flux.

\subsection{Sensitivity Limits}
\label{sec:SensitivityLimits}

The sensitivity limits for Xenon1T were calculated assuming the Xenon1T efficiency and energy window~\cite{XENON:2018voc} and negligible background in the CE$\nu$NS channel. In that case, the 90\% C.L. limit can be calculated by looking for the Poisson distribution $P$ expectation value $\lambda$ that gives a 90\% probability of observing at least one event. That value can be calculated as $1-P(\lambda, k=0) = 0.9$, which yields $\lambda \approx 2.3$ events.

To calculate the sensitivity limits for the DARWIN-like experiments, we consider two cases. The more conservative case (1) assumes the Xenon1T efficiency and energy window~\cite{XENON:2018voc} with 40 tonne-year exposure. In case (2), we assume 100\% detector efficiency, the nuclear recoil energy window extending down to 1~keV, and 200$\;$tonne$\;$year exposure. 

For the larger exposures, as expected in DARWIN, we calculate the 90\% C.L. sensitivity limits using test statistics that assume Wilk's theorem~\cite{Wilks:1938dza}, i.e., that the test statistics follows the chi-squared distribution with 2 degrees of freedom and the likelihood functions are the Poisson distributions. We also include the marginalization over the normalization uncertainty with the one $\sigma$ uncertainties respectively for the atmospheric neutrino flux $\sigma_\mathrm{atm} = 25\%$ and the solar neutrino flux $\sigma_\mathrm{sol} = 10\%$ with the Gaussian pull terms. The final expression for the chi-square test statistic minimized over the $x=[-0.99, 5]$ and $y=[-0.99, 5]$ parameter ranges 
is then
\begin{widetext}
\begin{equation}
 \Delta \chi^2_\mathrm{{min} \{x,y\}}  =  -2 \left(\sum_i N_{sig}^i + x N_\mathrm{sol}^i + y N_\mathrm{atm}^i + N_\mathrm{obs}^i \left( \log\left(N_\mathrm{atm}^i + N_\mathrm{sol}^i\right) - \log\left(N_\mathrm{obs}^i\right) \right) \right) + \left(\frac{x}{\sigma_\mathrm{sol}} \right)^2 + \left(\frac{y}{\sigma_\mathrm{atm}} \right)^2  \ ,
\end{equation}
\end{widetext}
where the number of expected events in the i-th recoil energy bin is the sum of the neutrino events from the two irreducible background contributions, i.e., the solar $N_\mathrm{sol}^i$ and atmospheric neutrinos $N_\mathrm{atm}^i$, and the number of observed events $N_\mathrm{obs} =  N_\mathrm{sig}^i + (1+x)N_\mathrm{sol}^i + (1+y)N_\mathrm{atm}^i$, where  $N_\mathrm{sig}^i$ is number of events from the DM annihilation. The bin size in our analysis is taken to be 1~keV. 

Figure~\ref{Fig1} shows the calculated sensitivity limits for Xenon1T as a red region, the DARWIN case (1) as a dark green region, and DARWIN case (2) as a light green region together with the existing limits from SK~\cite{Olivares-DelCampo:2017feq} (gray solid line) and projected sensitivities for JUNO~\cite{JUNO:2023vyz} and HK~\cite{Olivares-DelCampo:2018pdl}.
While the limits from Xenon1T are not competitive with the limits from the ordinary neutrino detectors, DARWIN has a chance to discover the DM annihilation signal in the region of the parameter space that JUNO and HK can also probe. Such a coincidence signal would strongly indicate DM annihilation to neutrinos.

\section{Discussion and Conclusions}
\label{sec:Discussion_and_Conclusions}

The increasing volumes and energy windows of the direct dark matter detectors will allow them to be excellent neutrino detectors thanks to the large cross sections for coherent nucleus-neutrino scattering~\cite{Freedman:1973yd, COHERENT:2017ipa}. We show that they can also be valuable low-mass dark matter detectors via indirect detection of the dark matter annihilation products -- neutrinos. 

In this work, we calculate the sensitivity limits on the MeV-mass dark matter annihilation cross section to neutrinos using the existing and planned large-scale direct detection experiments.
If the dark matter annihilation channel is exclusively into the third mass eigenstate, we find that planned direct dark matter detectors such as DARWIN are competitive with the current best limits coming from the Super-Kamiokande neutrino telescope. In addition, we find that future large-scale neutrino detectors, such as JUNO~\cite{JUNO:2023vyz}, Hyper-Kamiokande~\cite{Bell:2020rkw, Bell:2022ycf}, or DUNE~\cite{Klop:2018ltd, Arguelles:2019ouk, Buckley:2022btu}, can probe the region of the mass and cross section for dark matter annihilation to the third mass eigenstate that overlaps with the projected sensitivity region for DARWIN-like experiments. If the annihilation signature is detected by both of these types of detectors, that could strongly indicate the possibility of dark matter annihilating to neutrino mass states.

We find that our sensitivity limits for DARWIN are consistent with the ones presented in Ref.~\cite{McKeen:2018pbb}. Our constraints are slightly less stringent, most likely due to our inclusion of the marginalization over the normalization uncertainties in the irreducible solar and atmospheric neutrino backgrounds in our sensitivity analysis or our choice of a more conservative values for the Galactic dark matter density in the solar circle and distance to the Galactic center than in Ref.~\cite{McKeen:2018pbb}.    

Here, we estimate the sensitivity limits using the dark matter annihilation in the Milky Way. However, the contribution from the diffuse neutrino background~\cite{Beacom:2006tt, Klop:2018ltd, Bell:2022ycf} from annihilation in the extragalactic sources would not change the overall picture of the coincidence detection. Adding that contribution could only improve the sensitivity limits for both large-scale CE$\nu$NS and neutrino detectors. The predictions indicate that the extragalactic contribution to neutrinos created by the dark matter annihilation should be anywhere within an order of magnitude to comparable to the Galactic contribution~\cite{Beacom:2006tt, Klop:2018ltd, Bell:2022ycf}.

It is surprising that the neutrino detection capabilities of these large-scale dark matter detectors might shed light on the properties of dark matter.

\begin{acknowledgments}
\textbf{Acknowledgments.---}
We are grateful for helpful discussions with Anupam Ray and Thomas Wong.
This work was supported in part by National Science Foundation grant PHY-2209578 at UCSD, by National Science Foundation grant No.\ PHY-2020275: \emph{Network for Neutrinos, Nuclear Astrophysics, and Symmetries} (N3AS) and by Department of Energy grant No.\ DE-AC02-07CH11359: \emph{Neutrino Theory Network Program}. 

\end{acknowledgments}

\phantom{i}

\bibliography{DM}

\end{document}